\documentclass[twoside,english, jstat]{iopart}
\pdfoutput=1
\usepackage[T1]{fontenc}
\usepackage[latin9]{inputenc}
\usepackage{geometry}
\usepackage{amsbsy}
\usepackage{amstext}
\usepackage{amsthm}
\usepackage{graphicx}
\usepackage[numbers]{natbib}

\makeatletter
\usepackage{iopams}
\theoremstyle{plain}
\newtheorem{thm}{\protect\theoremname}

\newcommand{\eqref}[1]{(\ref{#1})}

\makeatother

\usepackage{babel}
\providecommand{\theoremname}{Theorem}

\begin{document}
\title[LLN in RPM]{Laws of large numbers in the Raise and Peel model}
\author{{\Large{}A.M. Povolotsky}}
\address{{\large{}Bogoliubov Laboratory of Theoretical Physics, Joint Institute
for Nuclear Research, 141980, Dubna, Russia}\\
{\large{}National Research University Higher School of Economics,
20 Myasnitskaya, 101000, Moscow, Russia}}
\ead{}
\begin{abstract}
We establish the exact laws of large numbers for two time additive
quantities in the Raise and Peel model, the number of tiles removed
by avalanches and the number of global avalanches happened by given
time. The validity of conjectures for the related stationary state
correlation functions then follow. The proof is based on the technique
of Baxter's T-Q equation applied to the associated XXZ chain and on
its solution at $\Delta=-1/2$ obtained by Fridkin, Stroganov and
Zagier.
\end{abstract}
\maketitle

\section{Introduction}

The Raise and Peel model (RPM) \citep{GNPR} is a non-equilibrium
model of fluctuating interface defined as a continuous time Markov
process. Its generator is given by a stochastic version of the Temperley-Lieb
(TL) Hamiltonian \citep{TL}, which in turn is a highly anisotropic
limit of the transfer matrix of O(1) dense loop model. The general
O(n) loop model has been intensively studied in context of theory
of critical phenomena since it was introduced in \citep{DMNS} as
a graphical lattice model expected to share the universality class
with O(n) vector model. The interest to the particular case of O(1)
dense loop model increased significantly after an observation by Razumov
and Stroganov of a nice combinatorial structure in the ground state
of an incarnation of the TL Hamiltonian, the XXZ quantum chain at
a specific combinatorial point \citep{RS-01}. The further studies
led to a discovery of the celebrated Razumov-Stroganov conjecture
on a miraculous connection between the ground state of the O(1) dense
loop transfer matrix and the set of configurations of the fully packed
loop model, in turn related to the alternating sign matrices \citep{RS-04}.
Also a number of conjectures on relations between RPM and the six
vertex model, XXZ model, the fully packed loop model, the O(1)
dense loop model and alternating sign matrices came from the studies
of this subject. Some of these conjectures, including the Razumov-Stroganov
conjecture itself \citep{CS}, were proven and much more are still
waiting for the proof (see \citep{AR-07,G} and references therein).

Many statements concern the structure of the ground state vector of
the transfer matrices or Hamiltonians under consideration. The term
combinatorial point refers to the fact that at specific values of
the parameters the coordinates of the ground state vector in a suitable
 basis can be normalized to be positive integers. Then the sums of
these integers turned out to be related to combinatorial numbers coming
from the enumeration of the alternating sign matrices. Treating the
suitably normalized coordinates as a probability distribution one
can also evaluate the marginal probabilities or the correlation functions
over this distributions. Basing on the analysis of finite size systems
several conjectures on exact ground state normalization and correlation
functions of the O(1) dense loop model with different types of boundary
conditions were proposed e.g. in \citep{RS-05,MNGB-1}.

In the present paper we concentrate on the proof of two such conjectures
in context of the RPM. Below we consider the RPM on the segment with
periodic boundary conditions. The model is formulated in terms of
1-d interface, which is the upper boundary of the densely packed set
of rhombic tiles placed onto a given substrate. It evolves by stochastic
addition of new tiles coming onto the interface from above. Depending
on the local shape of the interface at the place where the tile arrives,
it either stays (adsorption) or leaves (reflection) or causes a non-local
avalanche, which removes tiles from the current configuration (desorption).
We distinguish global and local avalanches depending on whether they
spread through the whole system or the do not respectively.

In the large time limit this stochastic process evolves towards the
stationary state. The vector of stationary probabilities of each configuration
is the eigenvector of the stochastic generator in the configurational
basis corresponding to the largest eigenvalue equal to zero. The eigenvector
components normalized to be the positive integers have nice combinatorial
properties mentioned above. Several conjectures of combinatorial nature
were proposed for the stationary state observables of RPM \citep{GNPR,AR-07}
and some other predictions on space and time correlation functions
were made with the use of conformal field theory \citep{AR-15,AR-13}.

Recently, the evolution of the RPM was also analysed beyond the stationary
state within the framework of the large deviation theory \citep{PPR}.
This theory allows one to study the statistics of the so called additive
functionals on the trajectories of Markov processes in the large time
limit. Two time-integrated quantities were studied: the number $\mathcal{N}_{t}^{\lozenge}$
of all tiles removed within the avalanches by given time $t$ and
the total number $\mathcal{N}_{t}^{\circlearrowleft}$ of global avalanches
happened by the same time. In particular a few rescaled cumulants
of these quantities were obtained asymptotically in the limit of large
system to the second and first order in the inverse system size respectively.
Note that in general the cumulants of time additive quantities depend
non-trivially on unequal time correlations and, thus, can not be obtained
from the stationary state observables. An exception is the first order
cumulants, the mean values, which are related to specific stationary
state observables, probabilities of events responsible for the change
of the quantities under consideration. Indeed, the limiting time averages
$\lim_{t\to\infty}t^{-1}\mathbb{E}\mathcal{N}_{t}^{\lozenge}$and
$\lim_{t\to\infty}t^{-1}\mathbb{E}\mathcal{N}_{t}^{\circlearrowleft}$
were related to stationary probabilities of specific interface configurations,
and were shown to match the leading orders of the previously conjectured
exact expressions of these probabilities. The aim of the present paper
is to establish the exact laws of large numbers for these quantities
at an arbitrary system size and hence to prove the related conjectures
for the stationary state correlation functions. Practically this means
the exact calculation of the above limits, which suggest a convergence
in expectation. Also, it is a simple fact from the theory of continuous-time
finite Markov chains that the convergence in expectation can be upgraded
to the almost sure convergence, which turns our statement to the strong
laws of large numbers.

Technically, calculations of the exact stationary state averages require
finding the stationary state probabilities explicitly, and, then,
summing over them with the quantity being averaged. This is not an
easy technical problem even on the level of finding the overall normalization
factor, see e.g. \citep{GBNM}. However in the case of the observables
related to the time-additive quantities the large deviation approach
allows one to reduce the calculations, to the analysis of the largest
eigenvalue of the deformed Markov generator rather than the eigenvector.
This deformation suggests including counting parameters into the matrix
of the generator, bringing the latter away from stochastic point.
The largest eigenvalue of such a deformed generator becomes the scaled
cumulant generating function of the time-additive quantities. The
scaled cumulants, aka the derivatives of this eigenvalue with respect
to the deformation parameters at zero (the stochastic point), can
be obtained using the connection of the RPM to the XXZ model. To this
end we use the technique based on solving the Baxter's T-Q equation
at the roots of unity developed in \citep{FSZ-1,FSZ-2}. Similar calculation
were done in \citep{S} for the nearest neighbor spin-spin correlation
functions in XXZ model with periodic boundary conditions. In our case
we make the calculations for the XXZ chain with twisted boundary conditions
corresponding to the stochastic generator, with the twist being one
of the counting deformation parameters. Note that in our case the
stationary state correlation functions  are defined in terms of the
interface configurations, rather than in spin basis natural for the
XXZ chain. The connection between these two bases is highly non-local
\citep{MDSA}, and in general the explicit relation between observables
in XXZ chain and the RPM is very non-obvious.

The paper is organized as follows. In Section 2 we define the model,
formulate the conjectures about two stationary state correlation functions
and state the main result as a Theorem \ref{thm:LLA}. We also discuss
there the relation of the stationary state conjectures to the law
of large numbers proved. In Section 3 we introduce the generator of
the Markov process and the deformed generator in terms an operator
in a finite dimensional ideal in the periodic Temperley-Lieb algebra
and using its R-matrix representation reduce the problem to finding
the largest eigenvalue of the XXZ chain with twisted boundary conditions.
It is reformulated in terms of solving the system of algebraic Bethe
ansatz equations. In the section 4 we rewrite the Bethe ansatz equations
in form of T-Q relation and use the method of its solution developed
in \citep{FSZ-1,FSZ-2} to calculate explicitly the derivatives of
the eigenvalue. The subsections 4.2 and 4.3 contain the technical
details of the calculation presented. We summarize the results and
discuss some perspectives in section 5.

\section{Raise and Peel model, problem statement and main result}

\subsection{Definition of the model}

\begin{figure}[h]
\centering{}\includegraphics[width=0.5\textwidth]{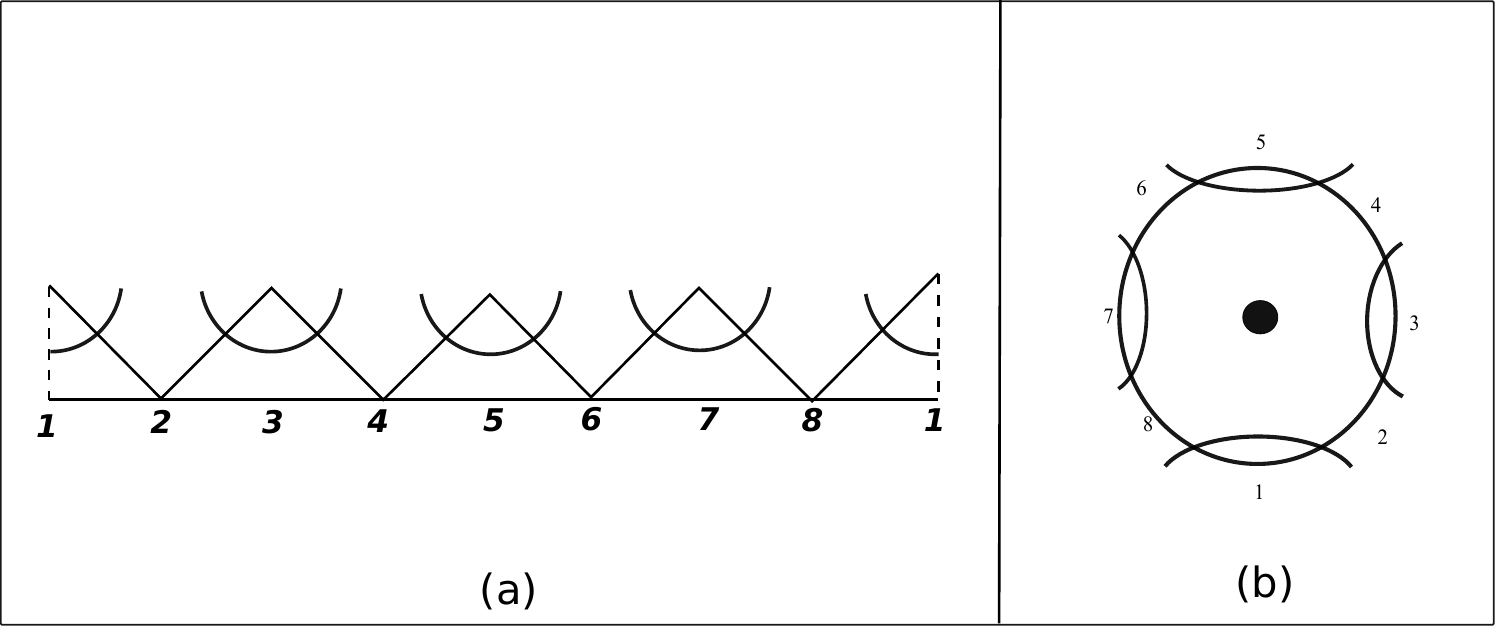}\caption{Substrate in PDP and link representation (a) and its projection onto
the punctured disk (b)\label{fig:substrate}}
\end{figure}

We consider the RPM on the segment of integer even length $L=2N$
with periodic boundary conditions. The evolution starts from the substrate,
which is the one-dimensional chain of rhombic tiles with the diagonals
of the length 2. The right end of the segment is identified with the
left end. The evolving random configuration is formed by adding tiles
on top of this substrate. The evolution preserves the tiles of the
substrate. This is why we picture only the upper halfs of tiles of
the substrate above the middle line, which we refer to as zero level,
see fig. \ref{fig:substrate}a.

An independent Poissonian clock is associated with each integer horizontal
position $1,\dots,L$. The clock rings after an exponentially distributed
waiting time with unit mean. When the clock at some position rings,
a tile falls onto the system at this position causing the change of
current configuration. The configuration of tiles established after
such a change, the stable configuration, is densely packed below its
upper boundary, the broken line, which makes steps either $(1,\sqrt{2})$
or $(1,-\sqrt{2})$, is forbidden to go down beyond the zero level
and should touch the first level at least once. Following to \citep{PPR}
we refer to it as the periodic Dyck path (PDP), and to such a system
representation as the PDP representation.

\begin{figure}
\centering{}\includegraphics[width=0.5\textwidth]{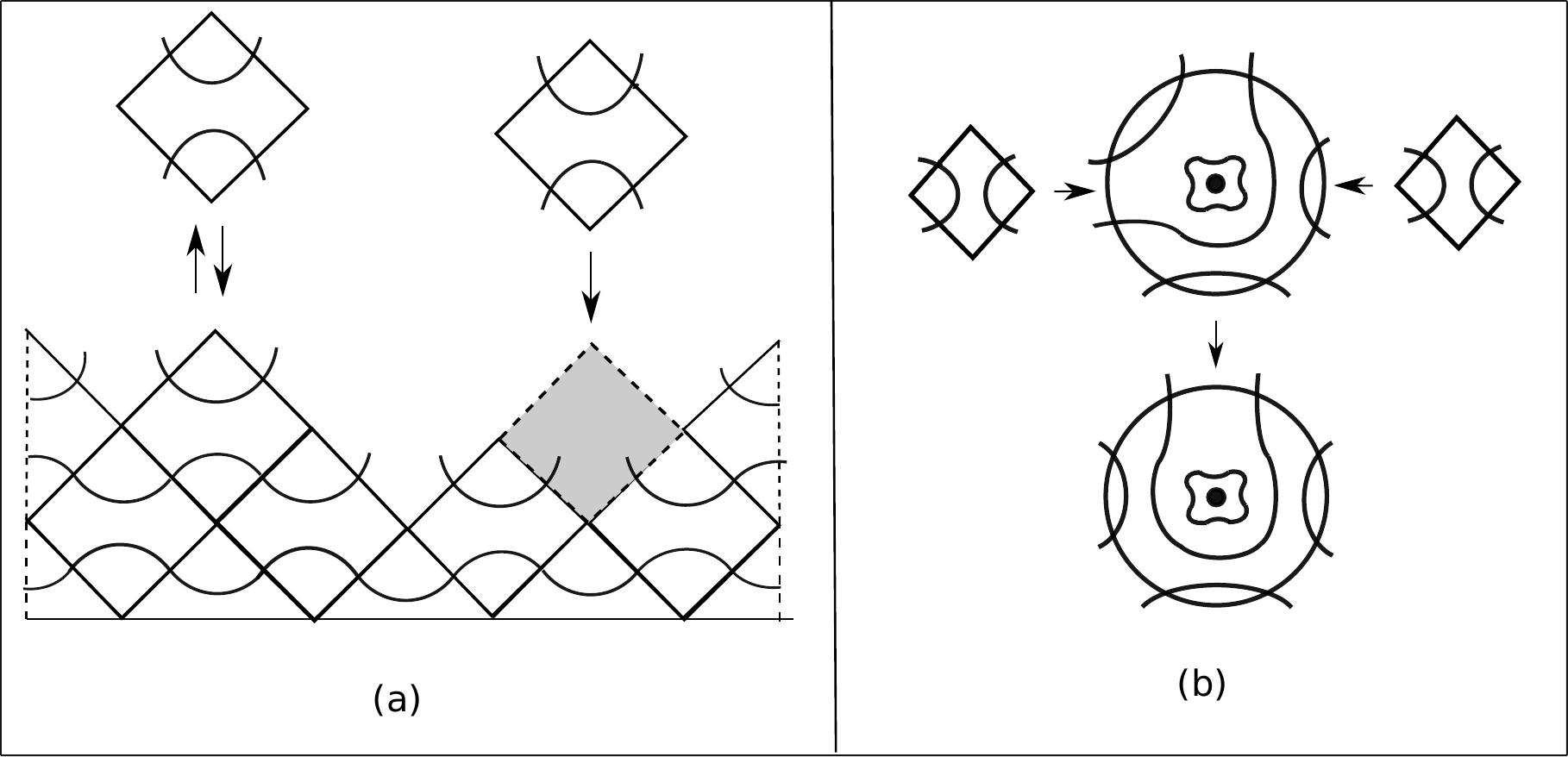}\caption{The events of reflection and adsorption. The left arriving tile disappears.
The right arriving tile stays at the shadowed place. (a) PDP and link
representation and (b) their projection onto the punctured disk .
\label{fig:reflection and adsorption}}
\end{figure}

\begin{figure}
\centering{}\includegraphics[width=0.5\textwidth]{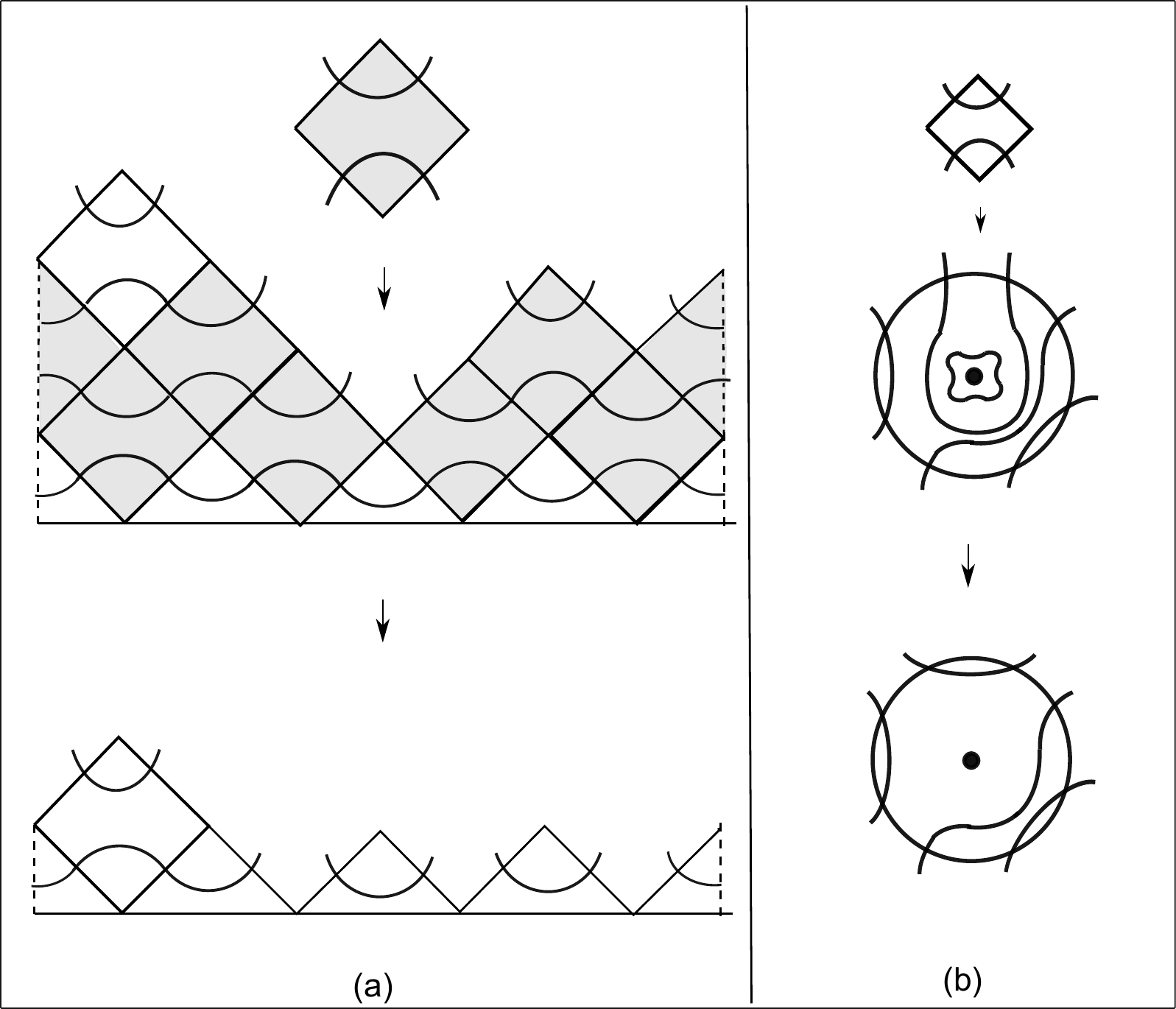}\caption{(a) The global avalanche caused by the tile completing two full layers
on top of the substrate removes the two full layers (shadowed tiles)
including the arrived tile. (b)The pair of non-contractible loops is
removed as soon as it appears due to addition of a pair of links.
\label{fig:global avalanche}}
\end{figure}

\begin{figure}
\centering{}\includegraphics[width=0.5\textwidth]{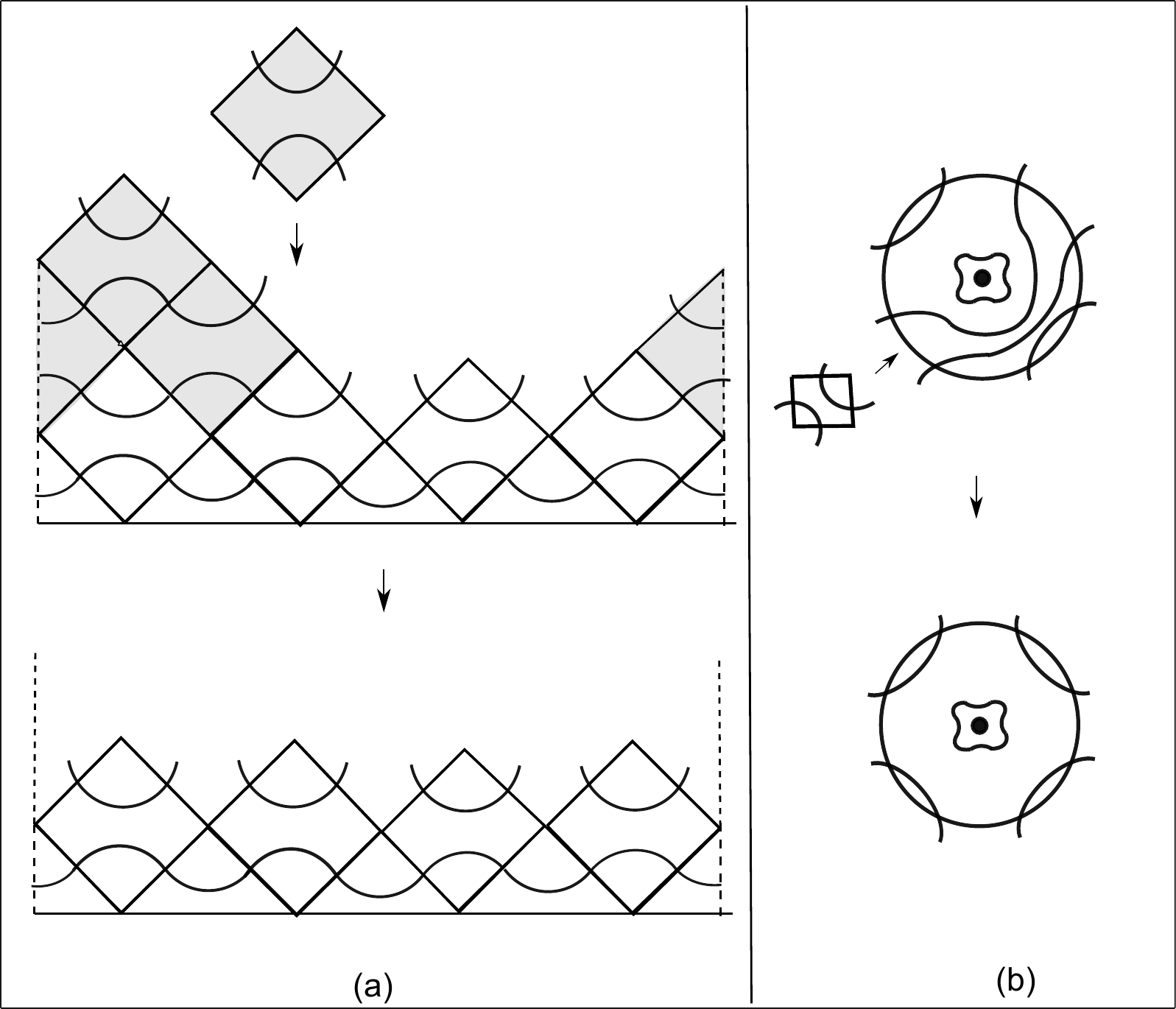}\caption{(a) The local avalanche removes the shadowed layer of tiles including
the arrived one. (b) An addition of a pair of links causes a rewiring
of free ends and loop contractions.\label{fig:local avalanche}}
\end{figure}

What happens, when the new tile arrives, depends on the local shape
of the interface at the arrival position. If the tile arrives at the
local maximum, the ``peak'', it gets reflected, i.e. simply removed,
see fig. \ref{fig:reflection and adsorption}a.
If the tile arrives at the local minimum, the ``valley'', it gets
absorbed, i.e. stays in the valley producing another stable configuration,
except when the tile has completed two full layers of tiles on top
of the substrate. In the latter case the configuration becomes unstable,
and the global avalanche happens, which removes the two full layers
of tiles including the arrived one, fig.\ref{fig:global avalanche}.
Arrival of the tile at the ``slope of a mountain'' also produces an
unstable configuration, that relaxes by the local avalanche that peels
off one layer of tiles at the level of arrival and upwards the mountain
up to the same level at the opposite slope including the tile arrived,
fig. \ref{fig:local avalanche}. Local and global avalanches represent
the non-local desorption events.

Another way to picture the process is to use the link representation.
To go from the PDP to the link representation we supply every tile
with a pair of links, one connecting between the bottom edges and
the other between the top edges of the tile. Then, every configuration
is represented by the non-crossing pairing of $N$ free ends of links
directed upwards plus may be one loop going around the whole system.
Arrival of new tiles causes rewiring of the free ends. The above rules
of the dynamics of tiles suggest that a contractible loop appeared
after the tile arrival must be contracted. As soon as a pair of non-contractible
loops has appeared, it should be removed.

This picture becomes more clear, when projected onto the disk punctured
in the center, see figs. \ref{fig:substrate}b-\ref{fig:local avalanche}b.
Then, a configuration is represented by either a directed non-crossing
pairing of $N$ points on the boundary of the disk or such a pairing
plus a non-contractible loop going around the hole in the center.
The term directed \citep{G} suggests that the hole allows one to
define a unique direction of the links. Then, the pairings of points
$(i,j)$ and $(j,i)$, i.e. where the link either goes around the
hole or not, are considered as different. As a new pair of links arrives,
it causes a rewiring of the links, where all the contractible loops
must be contracted. As a pair of non-contractible loops appears, they
should be removed.

The link representation traces back to the O(1) model, which is the
statistical physics model, defined as an ensemble of tilings of plain
domains with rhombic tiles with two possible orientations of pairs
of links connecting the adjacent edges. Each orientation of the tile
is assigned a statistical weight. In the stochastic version of such
a model the sum of the two weights is equal to one. The continuous
time limit suggests that the weight of one orientation is infinitesimally
small. The RPM in the PDP and in the link representation corresponds
to the continuous time stochastic O(1) tiling of the half-plain, were
we move through the tiling layer by layer, tracing only the connectivity
of the ends of the links at a given layer and the odd-counted non-contractible
loops. All the contractible loops fall out of  consideration as well
as the pairs of  non-contractible loops. More on this connection see
in \citep{GBNM}.

There is yet another representation of RPM obtained from identifying
every PDP with a configuration of $L/2$ particles on the one-dimensional
lattice with $L$ sites. To this end, one uses the usual interface-particle
system mapping. Every of steps $(1,\sqrt{2})$ or $(1,-\sqrt{2})$
of PDP is identified with particle or hole, respectively. The dynamics
of particles, which obey the exclusion interaction then follow from
the RPM dynamics. This system was introduced and analyzed in \citep{AR-13},
where it got a name ``Nonlocal asymmetric exclusion process'' NAEP
due to nonlocal particle jumps, corresponding to the avalanches in
the RPM.

\subsection{Stationary state and laws of large numbers for the avalanche currents.}

The above Markov process with finite state space having an irreducible
transition matrix evolves towards the unique stationary state. As
was discussed in the introduction, the vector of the stationary state
probabilities, i.e. the eigenvector of the transition matrix corresponding
to the non-degenerate zero eigenvalue, reveals nice combinatorial
properties. In particular, it can be normalized so that its smallest
component is one and the others are positive integers, whose sum is
given by the number of the half-turn symmetric alternating sign matrices.
Then many conjectures on correlation functions have simple rational
form. Two conjectures we address here are as follows.

\textbf{Conjecture 1.}The average number $n^{\triangle}$ of peaks
(local maxima) as well as valleys (local minima) in the stationary
state is
\begin{equation}
\mathbb{E}_{\mathbb{P}_{st}}n^{\triangle}=L\cdot\frac{3L^{2}}{8(L^{2}-1)}\label{eq:minimal 1-nests}
\end{equation}

\textbf{Conjecture 2.} Stable configurations which do not have valleys
on level $0$ and have a single valley on level $1$ appear in the
stationary state with probability
\begin{equation}
\mathbb{P}_{st}(\Omega_{\circlearrowleft})=\frac{3L}{4(L^{2}-1)},\label{eq:single nest}
\end{equation}
where we used the notation $\Omega_{\circlearrowleft}$ for the set
of such configurations.

The first conjecture for the number of peaks was proposed in \citep{GBNM},
where the subconfigurations corresponding to peaks were referred to
as minimal one-nests. It is worth noting that conjectures similar
to Conjecture 1 for RPM with different boundary conditions were proposed
in \citep{GNPR,APR}.

The second conjecture follows from the probability distribution of
the number of nests in a configuration conjectured in \citep{G}.
Specifically the probability of configuration with a single nest, obtained
from the conjectured expression, is twice larger than our value in
(\ref{eq:single nest}). Pronounced in the language of link representation,
the pairings considered in \citep{G} ignore the presence of non-contractible
loops. Thus, the configurations considered in \citep{G} correspond
to pairs of configurations in our system with and without non-contractible
loop having the same connectivity structure and equal probability.
Hence, the factor of $1/2$ separates those configurations with a
non-contractible loop present, where a second loop appears after an
addition of a single pair of links, like in fig. \ref{fig:global avalanche}b.

The above stationary state correlation functions are related to the
laws of large numbers for the avalanche currents $\mathcal{N}_{t}^{\lozenge}$
and $\mathcal{N}_{t}^{\circlearrowleft}$ introduced in \citep{PPR}.
Below we argue that the conjectures are equivalent to the following
theorem, which is the main result of this paper.
\begin{thm}
\label{thm:LLA}The following limits hold with probability one
\begin{equation}
\lim_{t\to\infty}\frac{\mathcal{N}_{t}^{\lozenge}}{t}=L\cdot\frac{5L^{2}-8}{8(L^{2}-1)},\label{eq: mean total}
\end{equation}
\begin{equation}
\lim_{t\to\infty}\frac{\mathcal{N}_{t}^{\circlearrowleft}}{t}=\frac{3}{4}\frac{L}{L^{2}-1}.\label{eq: mean global}
\end{equation}
\end{thm}

The value (\ref{eq: mean global}) for the stationary state average
current was conjectured by P.~Pyatov basing of the direct evaluation
of the stationary state eigenvector of the finite size transition
matrices, and then applied to predict the mean current in the NAEP
\citep{AR-13}. Both formulas were confirmed asymptotically in the
leading orders in $1/L$ in \citep{PPR}.

To show the first connection let us introduce two other time integrated
quantities $\mathcal{N}_{t},\mathcal{N}_{t}^{\triangle}$- the total
number of tiles arrived at the system and the number of tiles arrived
at the peaks by the time $t$. Then, we note that the total number
of tiles arrived by the time $t$ is the number of tiles on top of
the substrate $n_{t}$ at time $t$ plus the tiles reflected by the
peaks and the tiles removed by avalanches, i.e
\[
\mathcal{N}_{t}=\mathcal{N}_{t}^{\triangle}+\mathcal{N}_{t}^{\lozenge}+n_{t}.
\]
We now note that $\mathcal{N}_{t}=N(Lt)$ is the usual Poisson process
with the arrival rate $L$. By the law of large numbers for the Poisson
process, see e.g. \citep{Kingman}, the limit
\[
\lim_{t\to\infty}\frac{\mathcal{N}_{t}}{t}=L
\]
holds almost surely, i.e. with probability one.

The process $\mathcal{N}_{t}^{\triangle}$ is an example of the so
called doubly stochastic Poisson process or Cox process \citep{C},
i.e. the Poisson process with random arrival rate $n_{t}^{\triangle}.$
One can think of it as of the Poisson process $N(\tau_{t})$ with
stochastic time $\tau_{t}=\int_{0}^{t}n_{t}^{\triangle}$. By ergodic
theorem for Markov processes, see e.g. \citep{GS}, the limit
\[
\lim_{t\to\infty}\frac{\tau_{t}}{t}=\lim_{t\to\infty}t^{-1}\int_{0}^{t}n_{t}^{\triangle}dt=\mathbb{E}_{\mathbb{P}_{st}}n^{\triangle}
\]
 holds almost surely. Hence the law of large numbers for $\mathcal{N}_{t}^{\triangle}$
also takes place \citep{Serfozo}
\[
\lim_{t\to\infty}\frac{\mathcal{N}_{t}^{\triangle}}{t}=\mathbb{E}_{\mathbb{P}_{st}}n^{\triangle}
\]
with almost sure convergence. Finally, since $n_{t}$ is bounded as
$t\to\infty$, the limit
\[
\lim_{t\to\infty}\frac{\mathcal{N}_{t}^{\lozenge}}{t}=L-\mathbb{E}_{\mathbb{P}_{st}}n^{\triangle}
\]
holds with probability one.

For the second conjecture we note that it gives exactly the probability
of a configuration, where an addition of a single tile completes two
full layers of tiles on top of the substrate. Thus $\mathcal{N}_{t}^{\circlearrowleft}$
is the doubly stochastic Poisson process with the random arrival
rate $\mathbb{I}_{C(t)\in\Omega_{\circlearrowleft}}$, where $\mathbb{\mathbb{I}}_{C\in\Omega_{\circlearrowleft}}$
is an indicator function of the set $\Omega_{\circlearrowleft}$ and
$C(t)$ is a random configuration at the time $t$. Since by ergodicity
\[
\lim_{t\to\infty}t^{-1}\int_{0}^{t}\mathbb{I}_{C(t)\in\Omega_{\circlearrowleft}}dt=\mathbb{E}_{\mathbb{P}_{st}}\mathbb{I}_{C\in\Omega_{\circlearrowleft}}=\mathbb{P}_{st}\left(\Omega_{\circlearrowleft}\right)\quad a.s.
\]
the law of large numbers also holds for $\mathcal{N}_{t}^{\circlearrowleft}$
\[
\lim_{t\to\infty}\frac{\mathcal{N}_{t}^{\circlearrowleft}}{t}=\mathbb{P}_{st}(\Omega_{\circlearrowleft})\quad a.s.
\]

In the next sections we find the limits for convergence in expectation
using the large deviation technique. As the almost sure convergence
to a constant implies the convergence in expectation to the same limit,
this proves the theorem \ref{thm:LLA} as well as the conjectures
1,2.

\section{RPM, XXZ and the Bethe ansatz}

Consider the periodic Temperley-Lieb algebra on $PTL_{L}$ \citep{L}
generated by $L$ operators $e_{1},\dots,e_{L}$ satisfying the usual
TL relations
\begin{eqnarray}
e_{i}^{2} & = & 2_{q}e_{i},\label{eq: TL1}\\
e_{i}e_{i\pm1}e_{i} & = & e_{i},\label{eq: TL2}\\
e_{i}e_{j} & = & e_{j}e_{i},\quad\left|i-j\right|>1,\label{eq: TL3}
\end{eqnarray}
where $i,j=1,\dots,L\,\,(\mbox{mod}\,\,L)$, with the periodicity
condition $e_{L+1}\equiv e_{1}$. The coefficient in (\ref{eq: TL1})
is the algebra coefficient $2_{q}=q+q^{-1}$ parameterized by $q\in\mathbb{C}\setminus\{0\}$.
We need its quotient algebra by relations
\begin{equation}
J_{L}I_{L}J_{L}=\kappa J_{L},\qquad I_{L}J_{L}I_{L}=\kappa I_{L},\label{eq: TL_periodic}
\end{equation}
where
\begin{equation}
I_{L}=e_{2}e_{4}\dots e_{L},\qquad J_{L}=e_{1}e_{3}\dots e_{L-1}\label{eq: I,J}
\end{equation}
are the (unnormalized) projectors, and $\kappa\in\mathbb{C}\setminus\{0\}$
is an algebra parameter. Finally we consider the finite dimensional
left ideal $\mathcal{J}_{L}$ generated by the monomial $J_{L}$.
The monomials from this ideal can be visualized as the stable configurations
of RPM read from left to right and from top to bottom, where the generator
$e_{i}$ corresponds to the tile at horizontal position $i$. Specifically,
$J_{L}$ represents the substrate. The other stable configurations
are obtained by multiplying $J_{L}$ by generators with subsequent
reduction of the monomials by successive application of the relations
(\ref{eq: TL1}-\ref{eq: TL_periodic}). It is easy to see that up
to the numerical factors, which are products of powers of $\text{\ensuremath{\kappa}}$
and $2_{q}$, the reduction is equivalent to the processes following
the tile arrival described above . These factors turn to units at
the stochastic point
\begin{equation}
\kappa=1,2_{q}=1.\label{eq:stochastic}
\end{equation}

Then, the in the basis of monomials generated by $J_{L}$ the matrix
of the operator
\[
\mathcal{L}^{*}=\sum_{i=1}^{L}(e_{i}-1)
\]
is stochastic. It is a forward generator of the RPM Markov process,
which describes the time evolution of the probability distribution
$\mathbb{P}_{t}(C)$ of system configuration $C$
\[
\partial_{t}\mathbb{P}_{t}(C)=\mathcal{L}^{*}\mathbb{P}_{t}(C):=\sum_{\{C'\}}(u_{C',C}\mathbb{P}_{t}(C')-u_{C,C'}\mathbb{P}_{t}(C)).
\]
Here $u_{C',C}$ is a transition rate from a configuration $C'$ to
$C$, which is either one or zero in the case of RPM.

Let us introduce a deformed generator depending on two parameters
$\alpha$ and $\beta$ with action
\[
\mathcal{L}_{\alpha,\beta}^{*}F(C)=\sum_{C'}\left(e^{\alpha\delta\mathcal{N}_{C,C'}^{\circlearrowleft}+\beta\delta\mathcal{N}_{C,C'}^{\lozenge}}u_{C',C}F(C')-u_{C,C'}F(C)\right),
\]
which is obtained from $\mathcal{L}^{*}$ by multiplying the off-diagonal
matrix elements by $e^{\alpha\delta\mathcal{N}_{C,C'}^{\circlearrowleft}+\beta\delta\mathcal{N}_{C,C'}^{\lozenge}},$
where $\delta\mathcal{N}_{C,C'}^{\circlearrowleft}$ and $\delta\mathcal{N}_{C,C'}^{\lozenge}$
are the increases of corresponding quantities $\mathcal{N}_{t}^{\circlearrowleft}$and
$\mathcal{N}_{t}^{\lozenge}$ within the transition from the $C'$
to $C$. Its matrix coincides with the matrix of the operator
\[
\mathcal{L}_{\alpha,\beta}^{*}=\sum_{i=1}^{L}(\tilde{e}_{i}-1)
\]
in the monomial basis with the rescaled operators
\[
\tilde{e}_{i}:=2_{q}^{-1}e_{i},\quad\tilde{I}_{L}:=\tilde{e}_{2}\tilde{e}_{4}\dots\tilde{e}_{L},\quad\tilde{J}_{L}:=\tilde{e}_{1}\tilde{e}_{3}\dots\tilde{e}_{L-1},
\]
where we set
\[
\kappa=e^{\alpha}\,\,\,\mathrm{and\,\,\,}2_{q}^{-1}=e^{\beta}.
\]

The importance of this operator is due to the fact that its largest
eigenvalue $\Lambda(\alpha,\beta)$ is the scaled joint cumulant generating
function of $\mathcal{N}_{t}^{\circlearrowleft}$and $\mathcal{N}_{t}^{\lozenge}$.
In particular
\begin{eqnarray*}
\left.\frac{\partial\Lambda(\alpha,0)}{\partial\alpha}\right|_{\alpha=0} & = & \lim_{t\to\infty}\mathbb{E}\frac{\mathcal{N}_{t}^{\circlearrowleft}}{t},\\
\left.\frac{\partial\Lambda(0,\beta)}{\partial\beta}\right|_{\beta=0} & = & \lim_{t\to\infty}\mathbb{E}\frac{\mathcal{N}_{t}^{\lozenge}}{t}.
\end{eqnarray*}
For further details see \citep{PPR}.

\subsection{Bethe ansatz}

The further solution for the largest eigenvalue is based on the use
of the R-matrix representation of the PTL$_{L}$ on the space $\mathcal{H}=(\mathbb{C}^{2})^{\otimes L}$
. In this space the generator $e_{i}$ can be represented by

\[
\!\!\!\!\!\!\!\!\!\!\!\!\!\!\!\!\!\!\!\!\!\!\!\!\!\!\!\!\!\!\!\!\!\!\!\!\rho(e_{i})=\left[u\sigma_{i}^{+}\otimes\sigma_{i+1}^{-}+u^{-1}\sigma_{i}^{-}\otimes\sigma_{i+1}^{+}-\frac{\cos\gamma}{2}(\sigma_{i}^{z}\otimes\sigma_{i+1}^{z}-\mathbf{1}\otimes\mathbf{1})+\frac{\mathrm{i}\sin\gamma}{2}(\sigma_{i}^{z}\otimes\mathbf{1}-\boldsymbol{1}\otimes\sigma_{i+1}^{z})\right],
\]
 where $\sigma_{i}^{+},\sigma_{i}^{-}$ and $\sigma_{i}^{z}$ are
the matrices

\[
\sigma^{+}=\left(\begin{array}{cc}
0 & 1\\
0 & 0
\end{array}\right),\,\,\,\sigma^{-}=\left(\begin{array}{cc}
0 & 0\\
1 & 0
\end{array}\right),\,\,\,\sigma^{z}=\left(\begin{array}{cc}
1 & 0\\
0 & -1
\end{array}\right)
\]
acting in the $i$-th copy of $\mathbb{C}^{2}$ and the parameters
$\gamma$ and $u$ are related to $q$ and $\kappa$ by

\begin{eqnarray}
e^{\mathrm{\boldsymbol{i}\gamma}} & = & q\,\,\,\,\,\,\cos\gamma=\frac{2_{q}}{2},\qquad\sin\gamma=\frac{q-q^{-1}}{2\mathrm{i}},\label{eq:gamma vs q}\\
\kappa & = & \left(u^{N}+u^{-N}\right)^{2}.
\end{eqnarray}
Then the representation of the deformed Markov generator
\[
\rho\left(\mathcal{L}_{(\alpha,\beta)}^{*}\right)=-e^{\beta}H_{XXZ}^{\Delta,u}-\frac{3L}{4}
\]
is expressed in terms of the Hamiltonian
\begin{equation}
H_{XXZ}^{\Delta,u}=-\sum_{i=1}^{L}\left[u\sigma_{i}^{+}\otimes\sigma_{i+1}^{-}+u^{-1}\sigma_{i}^{-}\otimes\sigma_{i+1}^{+}+\frac{\Delta}{2}\sigma_{i}^{z}\otimes\sigma_{i+1}^{z}\right]\label{H-XXZ}
\end{equation}
of the anti-ferromagnetic Heisenberg quantum chain of $L$ spins 1/2
with anisotropy $\Delta=-\cos\gamma$. Note that by a unitary transformation
the parameter $u$ can be transferred from the Hamiltonian to twisted
boundary conditions:
\[
\left\{ \sigma_{L+1}^{\pm}=u^{\pm L}\sigma_{1}^{\pm},\,\,\,\sigma_{L+1}^{z}=\sigma_{1}^{z}\right\} .
\]
Finally, the largest eigenvalue of the operator $\mathcal{L}_{(\alpha,\beta)}^{*}$
is
\begin{eqnarray}
\Lambda_{0}(\alpha,\beta) & = & -e^{\beta}E_{L}^{XXZ}(\Delta,u)-\frac{3L}{4},\label{eq: Lambda_0(alpha,beta)}
\end{eqnarray}
where $E_{L}^{XXZ}(\Delta,u)$ is the groundstate eigenvalue of $H_{XXZ}^{\Delta,u}$
found in the sector $S_{z}=0$ isomorphic to the ideal $\mathcal{J}_{L}.$

The eigenvalues of $H_{XXZ}^{\Delta,u}$ in the sector $S_{z}=L/2-N$,
where $-L/2\leq N\leq L/2$ are given in terms of solutions of the
system of $N$ Bethe ansatz equations (BAE)
\begin{equation}
z_{i}^{L}=\left(-1\right)^{N-1}\prod_{j=1}^{N}\frac{1-2u^{-1}\Delta z_{i}+u^{-2}z_{i}z_{j}}{1-2u^{-1}\Delta z_{j}+u^{-2}z_{i}z_{j}},\,\,\,i=1,\dots,N.\label{eq:BAE-z}
\end{equation}
Then, the energy corresponding to a particular solution is

\[
E_{XXZ}=\Delta\left(2N-\frac{L}{2}\right)-\sum_{i=1}^{m}\left[uz_{i}^{-1}+z_{i}u^{-1}\right].
\]
Making a variable change
\[
z_{i}=u\frac{x_{i}-q}{1-qx_{i}}
\]
we obtain

\begin{equation}
u^{L}\left(\frac{x_{i}-q}{1-qx_{i}}\right)^{L}=\left(-1\right)^{N-1}\prod_{j=1}^{N}\frac{q^{2}x_{j}-x_{i}}{q^{2}x_{i}-x_{j}},\quad i=1,\dots,N,\label{eq:BAE-x}
\end{equation}
and

\begin{eqnarray}
E_{XXZ} & = & -\sum_{i=1}^{N}\left[\frac{q-q^{-1}}{1-x_{i}}+\frac{1-q^{2}}{q-q^{-1}x_{i}}+\frac{3}{2}2_{q}\right].\label{eq:EXXZ}
\end{eqnarray}
As a result, for the rescaled cumulant generating function of avalanche
currents in RPM we obtain
\begin{eqnarray}
\Lambda_{0}(\alpha,\beta) & = & \frac{1-q^{2}}{1+q^{2}}\sum_{i=1}^{N}\left[\frac{1}{1-qx_{i}}-\frac{q}{q-x_{i}}\right]-L,\label{eq:Lambda}
\end{eqnarray}
where we should set $L=2N$. To proceed with the proof of the statements
of interest we need to identify the groundstate solution of the system
of BAE for $L=2N$ and to evaluate the derivatives of $\Lambda_{0}(\alpha,\beta)$
with respect to $\alpha$ and $\beta$ (i.e. with respect to $q$
and $u$) at the stochastic point $\alpha=\beta=0,$ corresponding
to
\[
q=e^{\mathrm{i}\pi/3},u=e^{\mathrm{2}\pi\mathrm{i}/(3L)}.
\]
The necessary technique based on the T-Q equation was developed by
Fridkin, Stroganov, Zagier (FSZ) in \citep{FSZ-1,FSZ-2}.

\section{Bethe ansatz equations as T-Q relation}

The system of $N$ BAE can be rewritten as a single functional relation
for the polynomial
\[
Q(x)=\prod_{i=1}^{N}(x-x_{i}),
\]
with the roots being the root of a particular solution of BAE.

Then the BAE then read as
\begin{equation}
T(x)Q(x)=u^{L/2}\phi\left(xq^{-1}\right)Q(xq^{2})+u^{-L/2}\phi\left(xq\right)Q(xq^{-2}).\label{eq:T-Q-1}
\end{equation}
Where we introduce a notation
\begin{equation}
\phi(x)=\left(1-x\right)^{L}.\label{eq:phi}
\end{equation}
This T-Q relation is obtained as a condition of divisibility of the
degree $(N+L)$ polynomial in the r.h.s. by $Q(x)$. The ratio is
the polynomial $T(x)$ of degree $L$, which is known to be the eigenvalue
of the transfer matrix of the corresponding asymmetric six-vertex
model \citep{Baxter}. The eigenvalues (\ref{eq:EXXZ}) of XXZ Hamiltonian
can be expressed in terms of either $Q(x)$ or $T(x)$ at specific
points, which yields the following formulas for $\Lambda_{0}(\alpha,\beta):$
\begin{eqnarray*}
\Lambda_{0}(\alpha,\beta) & = & \frac{1-q^{2}}{1+q^{2}}\left(q^{-1}\frac{Q'(q^{-1})}{Q(q^{-1})}-q\frac{Q'(q)}{Q(q)}\right)-L\\
 & = & \frac{1}{1+q^{2}}\left(q(1-q^{2})\left.\frac{d\ln\tilde{T}(x)}{dx}\right|_{x=q}-L\right).
\end{eqnarray*}
It is also worth noting that multiplying the Bethe equations, we obtain
\[
\left(\prod_{i=1}^{N}z_{i}\right)^{N}=1.
\]
 For every state we should choose one of $N$ roots of unity for the
product $\prod_{i=1}^{N}z_{i}.$ The groundstate Bethe vector is translationally
invariant, so that the corresponding solution satisfies to
\[
\prod_{i=1}^{N}z_{i}=1,
\]
 i.e.
\[
\frac{Q(q^{-1})}{Q(q)}=\left(-\frac{1}{q}\right)^{N},
\]
which, being used in $T-Q$ relation (\ref{eq:T-Q-1}) at $x=q$,
yields
\begin{equation}
T(q)=\frac{\left(1-q^{2}\right)^{L}Q(q^{-1})}{Q(q)}=\left(1-q^{2}\right)^{L}\left(-\frac{1}{q}\right)^{N}.\label{eq: T(q)}
\end{equation}

\subsection{FSZ solution. }

The formulas of the previous subsection were valid for arbitrary $L$
and $N$. From now on we imply that $L=2N$ and write everything in
terms of $N$ for simplicity. Since we use multiplicative parametrization
rather than the additive of \citep{FSZ-1,FSZ-2}, to be self-contained
we briefly sketch the arguments of those papers. Let us introduce
notations
\[
q_{i}=Q(q^{2i}x),\quad t_{i}=T(q^{2i}x),\quad\phi_{i}=\phi(q^{i}x),\quad i\in\mathbb{Z}.
\]
It follows from (\ref{eq:T-Q-1})

\[
q_{k}t_{k}=u^{N}\phi_{-1}q_{k-1}+u^{-N}\phi_{1}q_{k+1},\quad k\in\mathbb{Z}
\]
When $q=e^{\pi\mathrm{i}/3}$, these are periodic functions $t_{k}=t_{k+3},q_{k}=q_{k+3},\phi_{k}=\phi_{k+6}$
\begin{eqnarray*}
q_{0}t_{0} & = & u^{N}\phi_{-1}q_{1}+u^{-N}\phi_{1}q_{2}\\
q_{1}t_{1} & = & u^{N}\phi_{1}q_{2}+u^{-N}\phi_{3}q_{0}\\
q_{2}t_{2} & = & u^{N}\phi_{3}q_{0}+u^{-N}\phi_{5}q_{1}
\end{eqnarray*}
For this system to have a solution for $q_{0},\dots,q_{2},$ the rank
of the matrix

\[
M=\left(\begin{array}{ccc}
-t_{0} & u^{N}\phi_{-1} & u^{-N}\phi_{1}\\
u^{-N}\phi_{3} & -t_{1} & u^{N}\phi_{1}\\
u^{N}\phi_{3} & u^{-N}\phi_{5} & -t_{2}
\end{array}\right)
\]
should be less than 3. It was observed that at the stochastic point
$u=e^{\frac{\pi\mathrm{i}}{3N}},q=e^{\frac{\pi\mathrm{i}}{3}}$ there
is a unique solution for $t_{k}$ that makes the rank of the matrix
equal to one
\begin{equation}
t_{k}=\phi_{3+2k},\label{eq: t_k}
\end{equation}
that luckily corresponds to the groundstate solution of BAE of the
XXZ model with$E_{XXZ}=-3L/4.$ Substituting
\begin{equation}
T(x)=(1+x)^{2N}\label{eq: T(x) FSZ}
\end{equation}
we obtain a q-difference equation for $Q(x)$
\begin{equation}
-(1+x)^{2N}Q(x)+u^{N}\left(1-xq^{-1}\right)^{2N}Q(xq^{2})+u^{-N}\left(1-xq\right)^{2N}Q(xq^{-2})=0\label{eq:T-Q diff}
\end{equation}
To solve this equation we introduce function
\begin{equation}
f_{Q}(x)=(1+x)^{2N}Q(x),\label{eq:Q-F_Q}
\end{equation}
which is a polynomial of degree $3N,$
\begin{equation}
f_{Q}(x)=\sum_{l=0}^{3N}f_{Q,l}x^{l}.\label{eq:f(x) poly}
\end{equation}
In terms of this function the relation (\ref{eq:T-Q diff}) reads
as
\[
-f_{Q}(x)+u^{L/2}f_{Q}(q^{2}x)+u^{-L/2}f_{Q}(q^{-2}x)=0.
\]
This relation fixes $N$ out of $3N+1$ coefficients $f_{Q,3l-1}=0,$
$l=1,\ldots,N$. The remaining $2N+1$ coefficients can be found up
to overall normalization factor from the condition that $f(x)$ has
a zero of order $2N$ at $x=-1.$ This fact is used as follows. We
require vanishing the derivatives of $f(x)$ up to order $(2N-1)$
at $x=-1$
\begin{eqnarray*}
f_{Q}^{(k)}(-1) & = & \sum_{n=k}^{3N}f_{Q,n}(-1)^{(n-k)}\frac{n!}{(n-k)!},\\
 & = & \sum_{l=-N}^{N}f_{Q,|3l+1|-1}(-1)^{-(|3l+1|-1-k)}\frac{(|3l+1|-1)!}{(|3l+1|-1-k)!}=0
\end{eqnarray*}
 for $k=0,\ldots,2N-1.$ As shown in \citep{FSZ-1,FSZ-2} this can
be satisfied by finding $2N+1$ numbers
\[
\beta_{l}=f_{Q,|3l+1|-1}(-1)^{-(|3l+1|-1)},\quad l=-L/2,\ldots,L/2.
\]
such that for any polynomial $P_{k}(x)$ of degree $k\leq(2N-1)$
the linear combination of its evaluations at points $\alpha_{l}=|3l+1|-1$
with these coefficients vanishes.

\[
\sum_{l=-N}^{N}\beta_{l}P_{k}(\alpha_{l})=0.
\]
In this way we define $\beta_{l}$ only up to a common factor
\[
\beta_{l}=c\prod_{s=-N}^{N,s\neq l}(\alpha_{l}-\alpha_{k}).
\]
Substituting this to $f_{Q}(x)$ we obtain

\begin{eqnarray*}
 &  & \!\!\!\!\!\!\!\!\!\!\!\!\!\!\!\!\!\!\!\!\!\!\!\!\!\!\!\!\!\!f_{Q}(x)=c\sum_{k=-N}^{N}(-1){}^{3N-(\left|3k+1\right|-1)}x^{3N-(\left|3k+1\right|-1)}\left(\prod_{\begin{array}{c}
s=-N,s\neq k\end{array}}^{N}\frac{1}{\left|3k+1\right|-\left|3s+1\right|}\right)\\
 &  & =N!\left(\frac{2}{3}-N\right){}_{N}\left(\sum_{k=0}^{N}\frac{x^{3k}}{(\frac{2}{3}-k)_{N}(N-k)!k!}+\sum_{k=0}^{N-1}\frac{x^{3k+2}}{(-k-\frac{2}{3})_{N+1}(N-k-1)!k!}\right),
\end{eqnarray*}
where
\[
(a)_{n}=a(a+1)\cdots(a+n-1)=\frac{\Gamma(a+n)}{\Gamma(a)}
\]
 is the Pochhammer symbol, and we choose the common factor $c$ such
that the largest degree coefficient is $f_{Q,3N}(x)=1.$ The latter
condition follows from our normalization of $Q(x)=x^{N}+O(x^{N-1})$
as $x\to\infty$.

One can see that the system of BAE is invariant with respect to the
simultaneous transformation $z_{i}\to1/z_{i}\,(\mathrm{or}\,x_{i}\to1/x_{i})$
and $u\to1/u.$ This suggests that for every solution of BAE $(x_{1},\dots,x_{N})$
there exists a solution $(x_{1}^{-1},\dots,x_{N}^{-1})$ of another
system obtained form the original one by the change $u\to1/u$, which
corresponds to a parity transformation. The analogue of $Q(x)$ in
that case will be
\begin{equation}
P(x)=x^{N}Q(1/x)Q^{-1}(0).\label{eq:P-Q}
\end{equation}

Such a polynomial solves the $T-P$ equation obtained from (\ref{eq:T-Q-1})
by the change $u\to1/u,$ and $T(x)\to x^{2N}T(1/x).$ The symmetry
$T(x)=x^{2N}T(1/x)$ corresponds to the simultaneous spin and space
reversal symmetry of the transfer matrix \citep{Baxter}, which in
turn suggests the equation for $P(x).$
\begin{equation}
T(x)P(x)=u^{-N}\phi\left(xq^{-1}\right)P(xq^{2})+u^{N}\phi\left(xq\right)P(xq^{-2}).\label{eq:T-P}
\end{equation}
 Note that the further calculation does not explicitly use arguments
based on the interpretation of $T(x)$ as a transfer matrix eigenvalue.
What we need is the solution of (\ref{eq:T-P}) with the same $T(x)$
as in (\ref{eq:T-Q-1}). However, the argument is what ensures the
existence of such a solution.

Similarly to $Q(x)$, at the stochastic point the polynomial $P(x)$
can be found in the form
\begin{equation}
P(x)=\frac{f_{P}(x)}{\left(1+x\right)^{2N}},\label{eq:P-f_P}
\end{equation}
where
\[
\!\!\!\!\!\!\!\!\!\!\!\!\!\!\!\!\!\!\!\!\!\!\!\!\!\!\!\!\!\!\!f_{P}(x)=N!\left(\frac{2}{3}\right)_{N}\left(\sum_{k=0}^{N}\frac{x^{3k}}{\left(k-N+\frac{2}{3}\right)_{N}k!(N-k)!}+\sum_{k=0}^{N-1}\frac{x^{3k+1}}{\left(k-N+\frac{1}{3}\right)_{N+1}k!(N-k-1)!}\right)
\]
For arbitrary $q$ and $u$ the two polynomials satisfy a Wronskian-like
relation
\begin{equation}
\phi\left(x\right)=\frac{u^{N}Q(qx)P(q^{-1}x)-u^{-N}Q(q^{-1}x)P(qx)}{u^{N}-u^{-N}}\label{eq:phiPQ}
\end{equation}
and the polynomial $T(x)$ can be written in terms of them
\begin{equation}
T(x)=\frac{u^{2N}Q(q^{2}x)P(q^{-2}x)-u^{-2N}Q(q^{-2}x)P(q^{2}x)}{u^{N}-u^{-N}}.\label{eq:TPQ}
\end{equation}

\subsection{Calculating derivatives}

Lastly we need to evaluate the derivatives of $\Lambda_{0}(\alpha,\beta)$
at $\alpha=0$ and $\beta=0.$

\subsubsection{The first argument}

For the derivative in $\alpha$ we have
\begin{eqnarray}
\left.\frac{\partial\Lambda_{0}(\alpha,0)}{\partial\alpha}\right|_{\alpha=0} & = & \frac{q(1-q^{2})}{1+q^{2}}\frac{T'_{u}(q)}{T(q)}\left.\frac{du}{d\alpha}\right|_{\alpha=0}\label{eq:dLambda/dalpha}\\
 & =\frac{1}{\mathrm{i}\sqrt{3}L} & \frac{q\left(1-q^{2}\right)}{\left(q^{2}+1\right)(q+1)^{L}}\frac{T'_{u}(q)}{T(q)},\nonumber
\end{eqnarray}
where we suggest that $q=e^{\mathrm{i}\pi/3}$, the subscript $u$
means the partial derivative in $u$ at $u=e^{\pi\mathrm{i/(3N)}},$
we taken into account that $du/d\alpha|_{\alpha=0}=(4iN\sin\left(\pi/3\right))^{-1}$and
$T(q)=(q+1)^{2N}$ follows from (\ref{eq:phi},\ref{eq: t_k}). Therefore
we need to calculate $T'_{u}(q)$, differentiating in the argument
and in $u$ and the substituting $q=u^{N}=e^{\mathrm{i}\pi/3}.$ Calculating
the derivatives of (\ref{eq:phiPQ}) and (\ref{eq:TPQ}) at $x=q$
and $x=q^{-2}$ respectively we obtain
\begin{eqnarray}
T'_{u}(q) & = & -\frac{N}{u}\frac{q+q^{-1}}{q-q^{-1}}T'(q)+A+B,\label{eq:Tuq}\\
0=\phi'_{u}(q^{-2}) & = & -\frac{N}{u}\frac{q+q^{-1}}{q-q^{-1}}\phi'(q^{-2})-A+\frac{B}{2},\label{eq:phiuq}
\end{eqnarray}
where
\begin{eqnarray*}
\!\!\!\!\!\!\!\!\!\!\!\!\!\!\!\!\!\!\!\!\!\!\!\!\!\!\!\!\!A & = & \frac{\left(Q'(-1)P(q^{-1})\right)_{u}q^{-2}+\left(Q(-1)P'(q^{-1})\right)_{u}-q^{2}(Q'(q^{-1})P(-1))_{u}-(Q(q^{-1})P'(-1))_{u}}{q-q^{-1}}\\
\!\!\!\!\!\!\!\!\!\!\!\!\!\!\!\!\!\!\!\!\!\!\!\!\!\!\!\!\!B & = & \frac{L}{u}\frac{Q'(-1)P(q^{-1})q^{-2}+Q(-1)P'(q^{-1})+q^{2}Q'(q^{-1})P(-1)+Q(q^{-1})P'(-1)}{q-q^{-1}}.
\end{eqnarray*}
Eliminating $A$ between (\ref{eq:Tuq}) and (\ref{eq:phiuq}) we
obtain
\[
T'_{u}(q)=\frac{3}{2}L\left(\phi'(q^{-2})+\frac{2}{q^{2}-1}(q^{-1}Q'(-q)P(q^{-1})+qQ(q^{3})P'(q^{-1})\right).
\]
Thus, we have reduced the problem to calculation of the values of
$P(x)$ and $Q(x)$ and their first derivatives at $x=-1,q^{-1}.$

\subsubsection{The second argument.}

For the derivative in $\beta$ we have

\begin{eqnarray}
 & \left.\frac{\partial\Lambda_{0}(0,\beta)}{\partial\beta}\right|_{\beta=0} & =\frac{2L}{\left(q^{2}-1\right)\left(q^{2}+1\right)^{2}}+\frac{\left(q^{4}-1\right)T_{q}(q)T'(q)}{\left(q^{2}-1\right)\left(q^{2}+1\right)^{2}T^{2}(q)}\label{eq:dLambda/dbeta}\\
 &  & +\frac{\left(1-q^{-2}\right)\left(T''(q)+T_{q}'(q)\right)+\left(1+4q^{-2}-q^{2}\right)T'(q)}{\left(q^{2}-1\right)\left(q^{2}+1\right)^{2}T(q)}\nonumber
\end{eqnarray}
To calculate $T(q),T'(q),T''(q)$ we use (\ref{eq: T(x) FSZ}) and
for the $q-$derivative $T_{q}(q)$ we differentiate ($\ref{eq: T(q)}$)
in $q$. The difficult part is to find $T_{q}'(q)$. To this end we
differentiate (\ref{eq:TPQ}) with respect to $q$ which yields
\[
T'_{q}(q)=\frac{q^{2}\left(A_{T}+B_{T}\right)-q^{-2}(A_{T}+B_{T})_{Q\leftrightarrow P}}{q-q^{-1}},
\]
 where the subscript $Q\leftrightarrow P$ indicates that the expression
in the brackets is obtained by interchanging the polynomials $P(x)$
and $Q(x)$ and
\begin{eqnarray*}
 &  & \!\!\!\!\!\!\!\!\!\!A_{T}=q^{2}\left(Q'_{q}(-1)P(q^{-1})+Q'(-1)P_{q}(q^{-1})\right)+q^{-2}(Q_{q}(-1)P(q^{-1})+Q(-1)P_{q}'(q^{-1})),\\
 &  & \!\!\!\!\!\!\!\!\!\!B_{T}=2\left(qQ'(-1)P(q^{-1})+q^{-2}Q''(q^{3})P(q^{-1})+Q(-1)P'(q^{-1})+q^{-1}Q(-1)P''(q^{-1})\right).
\end{eqnarray*}

Then from (\ref{eq:phiPQ}) we obtain
\[
0=\phi_{q}'(q)=\frac{q(A_{\phi}+B_{\phi})-q^{-1}(A_{\phi}+B_{\phi})_{Q\leftrightarrow P}}{q-q^{-1}},
\]
where we find that
\[
A_{\phi}=-(A_{T})_{Q\leftrightarrow P},\quad2B_{\phi}=\left(B_{T}\right)_{Q\leftrightarrow P}.
\]
which finally yields
\begin{equation}
T'_{q}(q)=\frac{3}{2}\frac{q^{2}B_{T}-q^{-2}(B_{T})_{P\leftrightarrow Q}}{q-q^{-1}}.\label{eq:T'_q}
\end{equation}
Thus, we again have reduced the problem to  evaluation of the values
of $P(x)$ and $Q(x)$ and their derivatives up to the second order
at points $x=-1,q^{-1}.$

\subsection{Calculations with $P(x)$ and $Q(x).$}

Below we obtain the necessary values of $P(x)$ and $Q(x)$ and their
derivatives. After substituting them to the formulas (\ref{eq:dLambda/dalpha}-\ref{eq:T'_q})
and lengthy but straightforward algebra we obtain the following results
\[
\left.\frac{\partial\Lambda_{0}(\alpha,0)}{\partial\alpha}\right|_{\alpha=0}=N\cdot\frac{5N^{2}-2}{4N^{2}-1},\,\,\,\left.\frac{\partial\Lambda_{0}(0,\beta)}{\partial\beta}\right|_{\beta=0}=\frac{3}{2}\frac{N}{4N^{2}-1},
\]
which prove Theorem \ref{thm:LLA}.

The result of this section is as follows.

\begin{eqnarray*}
Q\left(q^{-1}\right) & = & \frac{(2N-1)!\Gamma(1/3-N)}{(N-1)!\Gamma\left(1/3\right)}\frac{\left(1-q^{-2}\right)}{\left(q^{-1}+1\right)^{2N}},\\
Q'\left(q^{-1}\right) & = & \frac{N(q-1)(N(3q-1)-q+1)}{(2N-1)\left(q-q^{-1}\right)}Q\left(q^{-1}\right)\\
Q''\left(q^{-1}\right) & = & -\frac{N(N(8(N-1)q-5N+7)+4(q-1))}{2(2N-1)\left(1+q^{-1}\right)\left(q-q^{-1}\right)}Q\left(q^{-1}\right)
\end{eqnarray*}

\begin{eqnarray*}
P\left(q^{-1}\right) & = & \frac{(2N-1)!\Gamma\left(2/3-N\right)}{(N-1)!\Gamma\left(2/3\right)}\frac{1}{\left(q^{-1}+1\right)^{2N-1}}\\
P'\left(q^{-1}\right) & = & \frac{N\left(\left(3N-2\right)\left(1-q^{-2}\right)-2(2N-1)\right)}{(2N-1)\left(q^{-1}+1\right)}P\left(q^{-1}\right),\\
P''\left(q^{-1}\right) & = & \frac{N\left(N^{2}(1-3q)^{2}-N(q+1)(9q-7)+2(q(q+2)-1)\right)}{2(2N-1)\left(q^{-1}+1\right)^{2}}P\left(q^{-1}\right)
\end{eqnarray*}

\begin{eqnarray*}
Q(-1) & = & \frac{3^{2N}\Gamma\left(2/3\right)N!}{(2N)!\Gamma\left(2/3-N\right)}\\
Q'(-1) & = & -\frac{N\left(N+1\right)}{(2N+1)}Q(-1)\\
Q''(-1) & = & \frac{N\left(N-1\right)\left(3N+4\right)}{6(2N+1)}Q(-1)
\end{eqnarray*}

\begin{eqnarray*}
P(-1) & = & \frac{3^{2N}N!\Gamma\left(1/3\right)}{(2N)!\Gamma\left(1/3-N\right)}\\
P'(-1) & = & -\frac{N^{2}}{2N+1}P(-1)\\
P''(-1) & = & \frac{N(N-1)(3N-2)}{6(2N+1)}P(-1)
\end{eqnarray*}

To evaluate the values and the derivatives of $P(x)$ and $Q(x)$
at $x=-1,q^{-1}$ we use their representation (\ref{eq:Q-F_Q},\ref{eq:P-f_P})
in terms of functions $f_{P}$ and $f_{Q}$.

I. At $x=q^{-1}$ we represent these functions in terms of the Gauss
hypergeometric functions
\begin{eqnarray*}
f_{Q}(x) & = & \frac{\Gamma(2/3)}{\Gamma(2/3-N)}\left(\frac{\Gamma(2/3)}{\Gamma(2/3+N)}\left._{2}F_{1}(-N,1/3-N,1/3;-x^{3})\right.\right.\\
 & + & \left.\frac{x^{2}N\Gamma(-2/3)}{\Gamma(1/3+N)}\left._{2}F_{1}(1-N,2/3-N,5/3;-x^{3})\right.\right),
\end{eqnarray*}
\begin{eqnarray*}
f_{P}(x) & = & \frac{\Gamma(2/3+N)}{\Gamma(2/3)}\left(\frac{\Gamma(2/3-N)}{\Gamma(2/3)}\left._{2}F_{1}(-N,2/3-N,2/3;-x^{3})\right.\right.\\
 & + & \left.\frac{xN\Gamma(1/3-N)}{\Gamma(4/3)}\left._{2}F_{1}(1-N,1/3-N,4/3;-x^{3})\right.\right).
\end{eqnarray*}
Then using the fact that $q^{3}=-1,$ the Chu-Vandermonde identity
$\left._{2}F_{1}(-n,a,c;1)\right.=(a+c)_{n}/(a)_{n}$ for evaluating
the hypergeometric function at the unit variable with $n$ being positive
integer, and the formula of a derivative of the hypergeometric function
$\partial F(a,b,c;x)/\partial x=abc^{-1}F(a+1,b+1,c+1;x)$ we obtain
stated formulas.

II. At $x=-1$ the same problem is more delicate as the functions
$f_{Q}(x)$ and $f_{P}(x)$ have zeros of multiplicity $L$ at this
point. Let us proceed with the calculation for $Q(x),$ which will
also give us the results for $P(x)$ with use of the relation (\ref{eq:P-Q}).
Calculating the derivatives of (\ref{eq:Q-F_Q}) and setting $x=-1$
we obtain
\begin{equation}
Q^{(k)}(-1)=\frac{k!}{(2N+k)!}f^{(2N+k)}(-1).\label{eq:Q^(k)}
\end{equation}
We need to evaluate $f^{(2N+k)}(-1)$ for $k=0,1,2.$ Let us introduce
the notation
\[
c_{N}=\frac{3^{2N}\Gamma\left(2/3\right)N!}{\Gamma\left(2/3-N\right)}.
\]
Now we prove the identities
\begin{eqnarray*}
\mathrm{a)} & f^{(2N)}(-1) & =c_{N},\\
\mathrm{b)} & f^{(2N+1)}(-1) & =-c_{N}N\left(N+1\right),\\
\mathrm{c)} & f^{(2N+2)}(-1) & =c_{N}\frac{N\left(N^{2}-1\right)\left(3N+4\right)}{6},
\end{eqnarray*}
which being substituted to (\ref{eq:Q^(k)}) yields the above stated
results. To prove the identities, we show that their l.h.s divided
by the r.h.s. is a difference of two parts, which satisfy the same
recurrent relations with respect to $N$. As their initial values
differ by one, the ratios are equal to one. To establish the recursion
we use the Maple implementation of Zeilberger's algorithm of finding
the recurrent relations for hypergeometric-like sums \citep{Zeilberger}.

a) For the first identity we have
\[
c_{N}^{-1}f^{(2N)}(-1)=a_{1}(N)-a_{2}(N),
\]
where
\begin{eqnarray*}
a_{1}(N) & = & 3^{-2N}\sum_{k=0}^{N}\frac{(-1)^{3k-2N}(3k)!}{(\frac{2}{3}-k)_{N}(N-k)!k!(3k-2N)!},\\
a_{2}(N) & =- & 3^{-2N}\sum_{k=0}^{N-1}\frac{(-1)^{3k+2-2N}(3k+2)!}{(-k-\frac{2}{3})_{N+1}(N-k-1)!k!(3k+2-2N)!}
\end{eqnarray*}
with convention $(-n)!=\infty$ for $n\in\mathbb{Z}_{>0}.$ The quantities
$a_{1}(n)$ and $a_{2}(n)$ satisfy the same recurrent relation
\[
\left(6+4\,n\right)a_{i}\left(n\right)+\left(-5\,n-8\right)a_{i}\left(n+1\right)+\left(n+2\right)a_{i}\left(n+2\right)=0,\quad i=1,2
\]
with initial conditions $a_{1}(1)=2,a_{1}(2)=5,a_{2}(1)=1,a_{2}(2)=4.$

b) For the second identity
\[
-\frac{f^{(2N+1)}(-1)}{c_{N}N(N+1)}=a_{1}^{(1)}(N)-a_{2}^{(1)}(N)
\]

\begin{eqnarray*}
a_{1}^{(1)}(N) & = & 3^{-2N}\sum_{k=0}^{N}\frac{(-1)^{3k-2N-1}(3k)!}{(\frac{2}{3}-k)_{N}(N-k)!k!(3k-2N-1)!N(N-1)},\\
a_{2}^{(1)}(N) & =- & 3^{-2N}\sum_{k=0}^{N-1}\frac{(-1)^{3k+1-2N}(3k+2)!}{(-k-\frac{2}{3})_{N+1}(N-k-1)!k!(3k+1-2N)!N(N+1)}
\end{eqnarray*}
The recursion for $a_{1}^{(1)}(n)$ and $a_{2}^{(1)}(n)$ is
\[
\left(6+4\,n\right)a_{i}^{(1)}\left(n\right)+\left(-5\,n-9\right)a_{i}^{(1)}\left(n+1\right)+\left(n+3\right)a_{i}^{(1)}\left(n+2\right)=0,\quad i=1,2
\]
with initial conditions $a_{1}^{(1)}(1)=1,a_{1}^{(1)}(2)=5/3,a_{2}^{(1)}(1)=0,a_{2}^{(1)}(2)=2/3.$

c)
\[
\frac{6f^{(2N+1)}(-1)}{c_{N}N\left(N^{2}-1\right)\left(3N+4\right)}=a_{1}^{(2)}(N)-a_{2}^{(2)}(N)
\]

\begin{eqnarray*}
a_{1}^{(2)}(N) & = & 3^{-2N}\sum_{k=0}^{N}\frac{6(-1)^{3k-2N-2}(3k)!}{(\frac{2}{3}-k)_{N}(N-k)!k!(3k-2N-2)!c_{N}N\left(N^{2}-1\right)\left(3N+4\right)},\\
a_{2}^{(2)}(N) & =- & 3^{-2N}\sum_{k=0}^{N-1}\frac{6(-1)^{3k-2N}(3k+2)!}{(-k-\frac{2}{3})_{N+1}(N-k-1)!k!(3k-2N)!N\left(N^{2}-1\right)\left(3N+4\right)}
\end{eqnarray*}
The recursion for $a_{1}^{(2)}(n)$ and $a_{2}^{(2)}(n)$ is
\begin{eqnarray*}
 &  & 2\left(2\,n+5\right)\left(3\,n+4\right)a_{i}^{(2)}\left(n\right)-5\left(3\,n+7\right)\left(n+2\right)a_{i}^{(2)}\left(n+1\right)\\
 &  & \quad\quad\quad\quad\quad\quad\quad\quad\quad\quad+\left(3\,n+10\right)\left(n+3\right)a_{i}^{(2)}\left(n+2\right)=0,\quad i=1,2
\end{eqnarray*}
with initial conditions $a_{1}^{(2)}(1)=1,a_{1}^{(2)}(2)=20/13,a_{2}^{(2)}(1)=0,a_{2}^{(2)}(2)=7/13.$

\section{Conclusion}

To summarize, we obtained the exact expressions for the avalanche
currents in the Raise and Peel and justified the previous conjectures.
This task was completed by using the large deviation arguments, which
reduce the calculation to evaluation of the derivatives of the largest
eigenvalue of the deformed generator of the process. Technically,
the eigenvalue of the transfer matrix both in spectral parameter and
in the twist parameter or in the quantum group parameter was calculated.
This work can be considered as the first step towards the construction
of the exact large deviation function of the avalanche currents, which
was obtained asymptotically for the large system size in \citep{PPR}.
In particular, the next step would be obtaining the second cumulants,
the diffusion coefficients and the mutual covariance of the two currents.
While their asymptotic expressions were obtained in \citep{PPR},
conjectures on exact expressions like those for the mean currents
are absent. Will the technique used here still be applicable in that
case is a question for further studies. Also it would be interesting
to find other currents, whose statistics could be studied in the same
way, e.g. the ones related to spectral parameters in nonuniform Bethe
ansatz.

\ack{}{}

We dedicate this work to Vladimir Rittenberg, our colleague and friend
who was always an inexhaustible source of questions, who inspired
us and motivated the work on this project. We are grateful to Pavel
Pyatov for useful discussions. The work is supported by Russian Foundation
of Basic Research under grant 17-51-12001 and by the HSE University
Basic Research Program jointly with Russian Academic Excellence Project
'5-100'.

\end{document}